\documentclass[12pt]{article}
\usepackage{euscript,amsmath,amssymb,epsfig,latexsym}

\newcommand{\be}{\begin{equation}}
\newcommand{\ee}{\end{equation}}
\newcommand{\bea}{\begin{eqnarray}}
\newcommand{\eea}{\end{eqnarray}}

\newcommand{\pd}{\partial}

\begin{document}
\title{\bf
\vspace{1cm}On the Null Energy Condition and Cosmology}
\author{
I. Ya. Aref'eva and I.V. Volovich\\ \,
 \\
{\it  Steklov Mathematical Institute}
\\ {\it Gubkin St.8, 119991 Moscow, Russia}}

\date {~}
\maketitle
\begin{abstract} Field theories which violate the null energy
condition (NEC) are of  interest for the solution of the cosmological
singularity problem and for models of cosmological dark energy with
the equation of state parameter $w<-1$. We discuss the consistency
of two
 recently proposed models that violate the NEC. The ghost
condensate model requires higher-order  derivative terms in the action. It leads to a
heavy ghost field and unbounded energy.  We estimate the rates of particles decay and
discuss possible mass limitations to protect stability of matter in the ghost condensate
model. The nonlocal stringy model that arises  from a cubic string field theory and
  exhibits a phantom behavior also
leads to unbounded energy. In this case the spectrum of energy is continuous and there
are no particle like excitations. This model admits a natural UV completion
since it comes from superstring theory.

\end{abstract}

\newpage
\section{Introduction}
There are general restrictions to the energy-momentum tensor
$T_{\mu\nu}$ of a physical system. Such restrictions are referred to
as energy conditions. They play an important role in general
relativity, in particular in considerations of the black holes and
cosmological singularities \cite{Hawking-Ellis,GV}
(see also \cite{0301273}-\cite{0606091} and
refs. therein).  A weak form of
the energy condition states that $T_{\mu\nu}n^{\mu}n^{\nu}\geq 0$
for any null vector $n^{\mu}$; it is called the null energy
condition (NEC).

The NEC is directly related with a restriction on the dark energy
equation of state parameter $w=p/\rho$.  The energy-momentum tensor
of dark energy  is $T_{\mu\nu}=diag(\rho,p,p,p)$ with positive
energy density $\rho$ and negative pressure $p$ \cite{DE1,DE2}. The
condition $w<-1$ implies violation of the NEC. $~$Since experimental
data allows \footnote{A direct search strategy to test the
inequality $w<-1$ has been proposed~\cite{0312430}.} a possibility
of $w<-1$   the study of such models attracts a lot of attention.
Models of bouncing from cosmological singularity also require
violation of the NEC \cite{GV,GVb}.

There are general results that coupled scalar-gravity models which
violate the NEC are unstable
\cite{0301273,0311312,0406043,0606091,0512260,0602178} but recently
there have been proposed the ghost condensate model \cite{0312099,0405054,0602178,Creminelli}
and the nonlocal stringy model \cite{0410443,calcagni,AJ,AK} which apparently lead
to consistent effective theories. Both these models include higher-order derivative terms.
In this note we analyze these two
models  using the Ostrogradski method.

The ghost condensate model requires higher-order derivative terms in
the action hence it leads to a  massive ghost field and unbounded
energy and as a result it could violate the stability of matter. We
estimate the rates of particles decay and discuss possible mass
limitations to protect stability of matter in the ghost condensate model.

The nonlocal stringy model that exhibits a phantom behavior also leads
to unbounded energy but in this case the spectrum of energy is
continuous and there are no particle like excitations. This model
admits a UV completion since it comes from the superstring theory.

In more details,
 the model is based on SFT formulation of a
fermionic NSR string with the GSO$-$ sector \cite{NPB}. In this
model a scalar field is the open string tachyon \cite{AJK}, which describes
according to the Sen conjecture~\cite{Sen-g} a dynamical transition
of a non-BPS D-brane to a stable vacuum (see \cite{review-sft} for
review). This stable string theory is supposed to be described by a
VSFT (Vacuum String Field Theory). The model that we are going to
present is an approximation to VSFT which is stable and has no
particle excitations corresponding to the open string.

A vector-scalar model violating NEC with excitations modes stable in
some region has been proposed in \cite{Rubakov}. There is  a
tensor-scalar dark energy model admitting phantom behavior at small
redshifts\cite{Starobinsky}. String-inspired and braneworld dark
energy models are also the subject of intensive study in the
last years (see for example \cite{0605701,0605265,0605039} and refs.
therein).

The paper is organized as follows. In Section 2 we discuss the ghost
condensate model. We show that the model actually includes two
non-relativistic scalar fields. One field is a massive ghost while
the other field has a positive energy. We evaluate the decay rate of
an ordinary particle in the ghost condensate model and obtain a
rather stringent limit to the mass of the particle if the lifetime
is greater then the Hubble time. In Section 3 we consider the
nonlocal stringy model. We used the Weierstrass product
representation and the Ostrogradski method to display the spectrum
of the model. We obtain that also for this model the energy is
unbounded from below but in this case there is no particle like
excitations.
\section{Ghost Condensate Model}
\subsection{Setup}

The ghost condensate model in Minkowski space (signature is
$(-,+,+,+)$) deals with the Lagrangian \cite{0312099,0405054,0602178,Creminelli}
 \begin{equation}
\label{EffL1} {\cal L} = M^4 P(X) + M^2 S_1(X) (\Box \phi)^2 + M^2
S_2(X) \pd^\mu \pd^\nu \phi \pd_\mu \pd_\nu \phi  \end{equation}
Here $\phi$ is a scalar field,
  \begin{equation} \label{EffL2}X = - \pd_\mu \phi \pd^{\mu}\phi=
  \dot{\phi}^2-(\nabla \phi)^2\end{equation}
and $M$ an arbitrary mass scale. Ghosts (phantoms) could come from
the term $P(X)$. There are higher-derivative terms in the Lagrangian
proportional to  $S_1$ and $S_2$ which also lead to ghosts as it
will be discussed below.

 If
\begin{equation}
\label{EffL3} P(X)=\frac{1}{8}(X-1)^2=-\frac{1}{4}\dot{\phi}^2+...
\end{equation}
then the kinetic term has the "wrong" sign and one has ghosts. This corresponds to the
"wrong" vacuum state solution of the field equations
\begin{equation}
\label{EffL4v} \phi_0=0
\end{equation}

It was proposed in \cite{0312099} that to cure the theory, by analogy with the Higgs mechanism,
one can consider the configurations with $P^\prime(X)=0$ as a candidate ground state.
There is a solution of the field equations of the form
\begin{equation}
\label{EffL4} \phi_0=t
\end{equation}
which satisfies $P^\prime(X)=0$.

If we consider small fluctuations $\pi(t,x)$ about this solution
\begin{equation}
\label{EffL5} \phi=t+\pi(t,x)
\end{equation}
then the Lagrangian (\ref{EffL3}) for quadratic fluctuations leads to the kinetic term
with the "proper" sign
\begin{equation}
\label{EffL6} {\cal L} = \frac{1}{2} M^4\dot{\pi}^2
\end{equation}
There is no the spatial kinetic term $ (\nabla \pi)^2$  in the quadratic Lagrangian for
$\pi$. The higher order terms proportional to $S_1$ and $S_2$ in (\ref{EffL1}) are added
to $P$ to get a spatial kinetic term. The following quadratic Lagrangian is considered in
\cite{0312099,0602178,0405054,Creminelli}
\begin{equation}
\label{EffL7} {\cal L} = \frac{1}{2} M^4\dot{\pi}^2 -
\frac{1}{2}\gamma^2 (\nabla^2 \pi)^2
\end{equation}
where $\gamma^2$ is a constant,
\begin{equation}
\label{EffL76}
 \gamma^2 = - 2 M^2 (S_1(1) + S_2(1))
\end{equation}

We note, however, that  the original Lagrangian (\ref{EffL1}) leads not to
(\ref{EffL7}) but to the following quadratic Lagrangian
\begin{equation}
\label{EffL8} {\cal L} = \frac{1}{2} M^4\dot{\pi}^2 -
\frac{1}{2}\gamma^2 (\Box \pi)^2
\end{equation}
where
$$
\Box \pi=-\ddot{\pi}+\nabla^2 \pi.
$$
The Lagrangian (\ref{EffL8}) includes the higher derivative term
 which leads to ghosts for all momenta. In fact it describes not a
 single but two scalar fields.
 To demonstrate
it let us consider theories with higher derivatives.

\subsection{Ostrogradski Method for Equations of Higher Order}

Let us consider  the Lagrangian
\begin{equation}
\label{EH1} {\cal L} =  \varphi \Box (\alpha+\beta\Box )\varphi
\end{equation}
where $\alpha $ and $\beta$ are real parameters.
By using the  Ostrogradski method \cite{Ost,Osr-review} we
introduce the fields
\begin{equation}
\label{EH2} \psi=(\alpha+\beta\Box )\varphi,~~\phi=\Box \varphi
\end{equation}
Then, modulo the surface terms, the Lagrangian (\ref{EH1}) can be written as
\begin{equation}
\label{EH3} {\cal L} = \frac{1}{\alpha} \psi \Box \psi - \frac{\beta}{\alpha}
\phi(\alpha+\beta\Box )\phi
\end{equation}
We see that independently of the signs of $\alpha$ and $\beta$ one of fields
$\phi$ and $\psi$ is a ghost. For instance if $\alpha>0$ and $\beta<0$ then we have
 the massless ordinary field $\psi$ and the  massive ghost $\phi$.

The Lagrangian
\begin{equation}
\label{EH4} {\cal L} = \varphi \Box^2\varphi
\end{equation}
is more complicated for studies. The limit $\alpha\to 0$ is singular and to compute it
one has first to make a canonical transformation \cite{PaisU,Volovich}.

\subsection{Ghosts in  the Ghost Condensate Model}

 Now let us discuss the Lagrangian (\ref{EffL8})
$$ {\cal L} =
\frac{1}{2} M^4\dot{\pi}^2 - \frac{1}{2}\gamma^2 (\Box \pi)^2
$$
Equation of motion is
\begin{equation}
\label{EH5} (\Box^2 +\mu^2\pd_t^2)\pi=0
\end{equation}
where $\mu^2=M^4/\gamma^2$ or
\begin{equation}
\label{EH6} (\pd_t^4-2\pd_t^2\nabla^2+\nabla^4+\mu^2\pd_t^2)\pi=0
\end{equation}
For the spatial Fourier transform of the field $\pi(t,k)$ the
equation (\ref{EH6}) reads
\begin{equation}
\label{EH7} (\pd_t^4+(2k^2+\mu^2)\pd_t^2+k^4)\pi=0
\end{equation}
The characteristic equation
\begin{equation}
\label{EH8} \omega^4-(2k^2+\mu^2)\omega^2+k^4=0
\end{equation}
has only real roots
\begin{equation}
\label{EH9}
\omega^2_{1,2}(k)=\frac{1}{2}\mu^2[1+2\frac{k^2}{\mu^2}\pm
\sqrt{1+4\frac{k^2}{\mu^2}}]
\end{equation}
and we obtain
\begin{equation}
\label{EH10} \pd_t^4+(2k^2+\mu^2)\pd_t^2+k^4=
(\pd_t^2+\omega^2_1)(\pd_t^2+\omega^2_2)
\end{equation}
Denote
\begin{equation}
\label{EH11} \sigma^2=\omega^2_{1}-\omega^2_{2}=\mu^2
\sqrt{1+4\frac{k^2}{\mu^2}}>0
\end{equation}
By using the Ostrogradski method we introduce the fields
\begin{equation}
\psi=(\pd_t^2+\omega^2_1)\frac{1}{\sigma}\pi,~~
\phi=(\pd_t^2+\omega^2_2)\frac{1}{\sigma}\pi
\end{equation}
and obtain that the Lagrangian (\ref{EffL8})
\begin{equation}
\label{EH13} {\cal L}=- \frac{1}{2}\gamma^2
\pi[\pd_t^4+(2k^2+\mu^2)\pd_t^2+k^4]\pi
\end{equation}
modulo the surface terms can be represented as
\begin{equation}
\label{EH14} {\cal L}=
\frac{1}{4}\gamma^2\phi(\pd_t^2+\omega^2_{1})\phi-
\frac{1}{4}\gamma^2\psi(\pd_t^2+\omega^2_{2})\psi
\end{equation}
We obtain two non-relativistic fields $\phi$ and $\psi$. If
$\gamma^2
>0$ then the field $\phi$ leads to ghosts for any momenta $k$.
Note that there is a mass gap for ghosts since $\omega_1(k)^2\geq
\mu^2.$ For large $k$ one gets a relativistic dispersion law
$\omega_1(k)\sim \sqrt{k^2+\mu^2}.$ The field $\psi$ does not have a
mass gap.

\subsection{Decay of Particles}

 Ghosts (phantoms) could have  negative energy.
 Therefore ordinary particles can decay into
 heavier particles
plus phantoms. This is discussed in \cite{0301273,0311312}.

 We want to
estimate the lifetime of the ordinary particle under this decay. We
do not have  a direct interaction describing this decay. What
we have  for sure is an interaction of the phantom with  gravity and
 interaction of gravity with the usual particles. A simple
diagram describing a decay of an ordinary particle $\psi$ into usual
particles $\psi$ and $\chi$ and two phantoms  is presented in Fig.1.
There $h$ is a graviton.

\begin{figure}[h!]
\centering
\includegraphics[width=5cm]{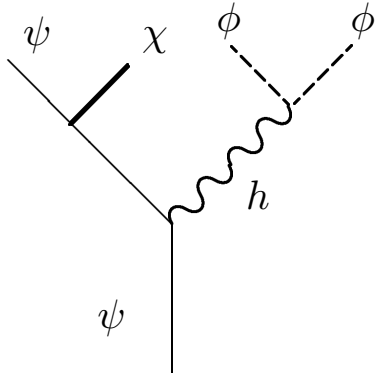}
  \caption{  Decay of  $\psi$ particle into two phantoms  and two
 other particles $\psi$ and $\chi$}
  \label{diagram}
\end{figure}

 We suppose that the interactions "gravity-phantom" and
 "gravity-ordinary fields" come from the mass terms
 in the action
 \begin{equation}
 S=M_p^2\int d^{4}x\sqrt{-g}R+\frac{1}{2}
\int d^{4}x\sqrt{-g}[(\pd\phi )^2+M^2\phi^2
\end{equation}
$$
 -(\pd\psi
)^2- m_\psi^2\psi^2-(\pd\psi )^2-
M_\chi^2\chi^2+\lambda_{\psi^2\chi} \psi^2\chi] $$ Here $\phi$ is a
phantom and $\psi$ and $\chi$ are ordinary fields. There is a cubic
interaction of the ordinary fields but they couple with the phantom
only through gravity. We assume that the mass $M_{\chi}$ of the
field $\chi$ is of the order of the large phantom mass $M$ and the
mass $m_{\psi}$ of the field $\psi$ is much smaller. From this
action we get the following interaction of gravity with matter
fields
 \begin{eqnarray}
 \label{ef1}
\sqrt{-g}\frac{M^2}{2}\phi^2\Rightarrow \frac{M^2}{M_p}\phi^2h\\
\label{ef2}\sqrt{-g}\frac{m_\psi^2}{2}\psi^2\Rightarrow \frac{m_\psi^2}{M_p}\psi^2h,
\end{eqnarray}
where $h$ is a scalar field  symbolically describing graviton.
Therefore, the coupling constants of gravity with the $\phi $ and
$\psi $ fields are \be \lambda_{h\phi^2}=\frac{M^2}{M_p},~~
 \lambda_{h\psi^2}=\frac{m_{\psi}^2}{M_p}\ee

 The decay rate of a particle to phantoms through
the channel (see Fig.1) \be \psi\to \psi+\chi+\phi_1+\phi_2 \ee is
\be \Gamma = \frac{1}{m_{\psi}}
    \int^{\Lambda}\frac{d^3p_{\phi_1}}{(2\pi)^32E_{\phi_1}}
        \frac{d^3p_{\phi_2}}{(2\pi)^32E_{\phi_2}}
        \frac{d^3p_{\psi}}{(2\pi)^32E_{\psi}}
        \frac{d^3p_{\chi}}{(2\pi)^32E_{\chi}}
        |{\mathcal M}|^2\ee

$$
(2\pi)^4
        \delta^{(4)}(p_{\psi_{in}} - p_{\phi_1}-p_{\phi_2}-p_{\psi}-p_{\chi}) \ ,
$$         where the
matrix element ${\mathcal M}$ is \be {\mathcal
M}=\lambda_{h\phi^2}\lambda_{h\psi^2}\lambda_{\chi\psi^2}
\frac{1}{p^2}\frac{1}{q^2+m_{\psi}^2} \ee Here $\Lambda$ is a
momentum cutoff, $p$ is the momenta of the internal graviton line
and $q$ of the internal $\psi$ line.
We assume that the dimensional  coupling constant
$\lambda_{\chi\psi^2}$ is of the order
$$\lambda_{\chi\psi^2}\sim m_{\psi}.$$

Using (\ref{ef1}) and
(\ref{ef2}) we get \be {\mathcal M}=
\frac{m_\psi^3M^2}{M_p^2}\frac{1}{p^2}\frac{1}{q^2+m_{\psi}^2}.
\ee
The integral is convergent even without cutoff. To estimate it we
can take
$$M\sim M_p\sim \Lambda$$

We estimate the decay rate as \be \label{five}\Gamma \sim
\frac{m_{\psi}^5}{M_p^5}M_p \ee The timescale for decay is $
\tau_\psi=1/\Gamma. $ A model will be phenomenologically viable if
the lifetime is greater than the Hubble time $H_0^{-1}$, i.e.
\be
\Gamma\sim \frac{m_{\psi}^5}{M_p^5}M_p <H_0\sim 10^{-60}M_p
\ee
Using $M_p\sim 10^{19}$GeV we get
\be
\label{five1}m_{\psi}<10^{7}\mbox{GeV}
\ee

A stringent limit to the mass one can obtain from the investigation
of the quartic interaction of the ordinary particles of the form
$\lambda \psi^3\chi$ with the coupling constant $\lambda$ of the
order one. Then instead of (\ref{five}) and (\ref{five1}) we get
\be
\label{three}\Gamma \sim \frac{m_{\psi}^3}{M_p^3}M_p \ee
and an uncomfortable result
\be
\label{three1}m_{\psi}<10^{-1} \mbox{GeV}.
\ee

\section{Nonlocal Stringy Model}
\subsection{Example of Field Equations of the Exponential Type}

Consider first the following action
\be
\label{Non1} {\cal L} =  \phi (e^{\Box}-1)\phi
 \ee
 The field equations are nonlocal since one can write them as integral equations.
 It serves as an example  for the nonlocal stringy Lagrangian which will be considered
 below. We don`t consider here the rigorous mathematical approach to
 such equations, see recent papers \cite{Vladimirov}.
By using the Weierstrass product
\be \label{Non2}
e^{z}-1=e^{z/2}z\prod_{j=1}^{\infty}(1+\frac{z^2}{\omega_j^2})
 \ee
 where
 $$\omega_j^2=4\pi^2 j^2$$
 we write
\be \label{Non3}
e^{\Box}-1=e^{\Box/2}\Box\prod_{j=1}^{\infty}(1+\frac{\Box^2}{\omega_j^2})
 \ee
Therefore \cite{PaisU} the Lagrangian (\ref{Non1}) has the same
spectrum as
\be
\label{Non4}
{\cal L} =  \psi_0 \Box\psi_0
+\sum_{j=1}^{\infty}\epsilon_j\psi_j(\Box^2+\omega_j^2)\psi_j
 \ee
Here \be \label{Non5} \epsilon_j=1/\omega_j^4 F^{\prime}(-\omega_j^2)
 \ee
where \be \label{Non6} F(z^2)=(e^{z}-1)\frac{e^{-z/2}}{z}=2\frac{\sinh z/2}{z}\ee

More general Lagrangian \be \label{Non7} {\cal L} =  \phi
(e^{\alpha\Box}-\lambda)\phi
 \ee
appears in p-adic string theory  (see \cite{p-adic,p-adic-book} and
refs therein). Here $\alpha$ and $\lambda$ are constants. The
spectrum of the theory can be read out of the Weierstrass product
\be \label{Non8}
e^{\alpha\Box}-\lambda=\lambda^{1/2}e^{\alpha\Box/2}
(\alpha\Box-\log\lambda)\prod_{j=1}^{\infty}
(1+\frac{(\alpha\Box-\log\lambda)^2}{\omega_j^2})
 \ee

\subsection{ Nonlocal Tachyon Field}

 An effective Lagrangian  for the tachyon field  coming from GSO$-$ sector of the fermionic NSR string
 which is described by cubic string field theory \cite{NPB}
 has the following form \cite{AJK}
\begin{equation}
{\cal L} =  \frac{1}{2}\phi e^{-\frac{1}{4}\Box}(\Box+m^2)\phi-\frac{\lambda}{4}\phi^4
\end{equation}

There is the "wrong" vacuum state solution of the field equations
\begin{equation}
\label{Nonv} \phi_0=0
\end{equation}
which leads to the tachyon equation for fluctuations about this vacuum:
\begin{equation}
\label{Nonv} (\Box+m^2)\phi=0
\end{equation}
There exists also another vacuum solution
\begin{equation}
\label{Nonv2} \phi_0=m/\sqrt{\lambda}
\end{equation}
One obtains the following Lagrangian for the small fluctuations about this solution:
\begin{equation}
\label{nca}
{\cal L} =  \frac{m^2}{2}\phi F(-\Box)\phi
\end{equation}
where
\begin{equation}
\label{Fforsft}F(z)= \left(-\xi^2z+1\right)e^{\frac{1}{4}z }-3,
\end{equation}
and we set $\xi^2=1/m^2.$

The characteristic equation
\begin{equation}
F(z)=0 \label{characteristic_m}
\end{equation}
has an infinite number of complex conjugated pairs of complex roots
$z=\kappa^2_j$ (see \cite{AK} for the consideration of these roots
in the case corresponding to the action (\ref{nca}) with
(\ref{Fforsft}) and \cite{AJV} for the case of arbitrary constants).
The roots $\kappa^2 _j$ are expressed by means of the Lambert
function $W(z)$ satisfying the equation $W(z)e^{W(z)}=z$.

We represent (\ref{Fforsft}) in terms of the Weierstrass product
\begin{equation}\label{prod}
F(z)=-2e^{\frac{1}{8}(4\xi^2-1)z}\prod_{j}(1-\frac{z}{\kappa ^2_j})
(1-\frac{z}{\kappa^{*2}_j})
\exp[z(\frac{1}{\kappa_j^2}+\frac{1}{\kappa_j^{*2}})]
\end{equation}

Now  we can write an appropriate  quadratic Lagrangian (up to total derivatives)
as
\be \label{new-fields} {\cal L}=\sum _{j}\epsilon
_j\psi_j\left(\Box -\kappa^2_j\right)\left(\Box
-\kappa^{*2}_j\right)\psi_j
\ee

Equations of motion with complex distinct frequencies were studied
in \cite{PaisU}.
Consider the following equation:
\be
\label{new-fields2} \left(\Box -\kappa^2\right)\left(\Box
-\kappa^{*2}\right)\phi=0 \ee where $\kappa=\nu+i\alpha.$ The
corresponding Hamiltonian is
 a linear superposition of pairs of complex conjugated oscillators
of the form: \be \label{new-fields3}
H=\frac{1}{2}\sum_k[P_1^2(k)+\omega(k)^2Q_1^2(k)+P_2^2(k)+\omega(k)^{*2}Q_2^2(k)]
\ee where
$$
P_1^*=P_2,~~Q_1^*=Q_2
$$
and
$$
\omega(k)^2=\kappa^2+k^2
$$
The spectrum of the Hamiltonian has the form
\be
\label{new-fields4}
n\nu(k)+\rho\alpha(k)
 \ee
 where we put $\omega(k)=\nu(k)+i\alpha(k).$ Here $n=0,\pm  1,...$
 while $\rho$ is continuous and ranges from $-\infty$ to $+\infty$.
 The energy spectrum is indefinite and continuous. Therefore no particle
 attributes can be ascribed to these modes \cite{PaisU}.

\section{Conclusions}

We have considered two recently proposed models which violate the NEC.
In both models there are higher derivative terms and we have used the
 Ostrogradski method
to study them. A straightforward application of this method to linear
theories indicates
that energies are unbounded.
Note that the problems with unbounded energy considered in this paper are mathematically similar to the
eigenvalue problem for the non positively defined hyperbolic
Klein-Gordon equation on Lorentzian manifolds
which is solved in  \cite{KozVol}.

One can expect that an incorporation of nonlinear
terms could drastically change the situation. In particular,
in nonlinear theories could exist
islands of  stability \cite{0301273,AKV,0605229,0612026}.

In the ghost condensate model with higher-order derivative terms it would be interesting  to find a UV
completion of the theory and study islands of stability.

 The nonlocal stringy model has been proposed to describe a decay of D-brane
 and though it also leads to unbounded energy  in
this case the spectrum of the  energy is continuous and there are no
particle like excitations. This model admits a UV completion since
it comes form  superstring theory but it would be very interesting to
find a direct mechanism of compensation of unbounded continuous spectrum.

\section*{Acknowledgements}

 The work of I.A. and I.V. is supported in part
by INTAS grant 03-51-6346. I.A. is also supported  by RFBR grant 05-01-00758  and Russian
President's grant NSh-2052.2003.1 and I.V. is also supported  by RFBR grant 05-01-00884  and Russian
President's grant NSh-1542.2003.1
\end{document}